\documentclass[aps,prd,superscriptaddress,nofootinbib,floatfix,amsmath,amssymb,onecolumn,12pt]{revtex4}

\usepackage{soul}

\usepackage[utf8]{inputenc}
\usepackage{amsmath}
\usepackage{slashed}
\usepackage{graphicx}
\usepackage[caption=false]{subfig}
\usepackage{xcolor}
\usepackage[pdftex]{hyperref}
 
\begin{document}

\title{Listening to Quantum Gravity?}
\author{Lawrence M. Krauss}
\affiliation{\scriptsize{The Origins Project Foundation, 4738 E. Rancho Dr, Phoenix AZ 85018 USA, lawrence@originsproject.org, corresponding author }}
	\author{Francesco Marino}
		\affiliation{\scriptsize{CNR-Istituto Nazionale di Ottica, Via Sansone 1, I-50019 Sesto Fiorentino (FI), Italy, francesco.marino@ino.cnr.it}}
\author{Samuel L. Braunstein}
\affiliation{\scriptsize{Computer Science, University of York, York Y010 5GH, United Kingdom, sam.braunstein@york.ac.uk}}
\author{Mir Faizal}
\affiliation{\scriptsize{Irving K. Barber School of Arts and Sciences,
University of British Columbia - Okanagan, Kelowna, British Columbia V1V 1V7, Canada, mirfaizalmir@googlemail.com}}
\affiliation{\scriptsize{Canadian Quantum Research Center 204-3002  32 Ave Vernon, BC V1T 2L7 Canada}}
	\author{Naveed A. Shah}
\affiliation{\scriptsize{Department of Physics, Jamia Millia Islamia, New Delhi - 110025, India, naveed179755@st.jmi.ac.in} }

\begin{abstract}

\vspace{0.7 in}
Recent experimental progresses in controlling classical and quantum fluids have made it possible to realize acoustic analogues of gravitational black holes, where a flowing fluid provides an effective spacetime on which sound waves propagate, demonstrating Hawking-like radiation and superradiance. We propose the exciting possibility that new hydrodynamic systems might provide insights to help resolve mysteries associated with quantum gravity, including the black hole information-loss paradox and the removal of spacetime singularities.

\vspace{0.3 in}
\centerline{\it{Essay written for the Gravity Research Foundation 2024 Awards for Essays on Gravitation.}}

\end{abstract}

\maketitle

\newpage

The last decade has witnessed the remarkable ability to probe what are otherwise unobservable gravitational effects in astrophysical phenomena using analogue systems that allow testing similar effects in a controlled terrestrial laboratory environment.  
Analogue gravity is based on the idea that, under appropriate conditions, collective excitations in condensed-matter systems that are induced by the medium in which they are propagating evolve just as quantum fields do in a curved spacetime  \cite{rev,barcelorev,facciorev}. The paradigmatic example is provided by sound waves (phonons) in an inhomogeneous flowing fluid \cite{rev,white,unruh,visser1,visser2}. The phonon velocity relative to the laboratory frame is modified by the flow, although phonons always travel at the speed of sound relative to the flowing fluid. 
The phonon trajectories in the spacetime (${\bf x}$,$t$) are implicitly defined by the quadratic equation 
\begin{equation}
-c_s^2 dt^2 + (d{\bf x} - {\bf v}dt)^2 = 0 \; ,
\label{cones}
\end{equation} 
where ${\bf v}$ is the velocity field of the fluid and $c_s$ is the local speed of sound. Since nothing in the fluid can propagate faster than phonons, Eq.~(\ref{cones}) defines the ``sound cones'' delimiting the region of causally connected events associated to each point in space and time. One can readily see that a spatially-inhomogeneous flow, ${\bf v}={\bf v}({\bf x})$, will tip the sound cones of a given event. As a result, the phonons experience an effective curved spacetime, the null geodesics of which are given by (\ref{cones}). The geometry can be specified by a Lorentzian metric tensor, the so-called acoustic metric, which has the general form
\begin{equation}
g_{\mu\nu} = \left(\begin{array}{cc}
  -c_s^2 + v^2  &  -{\bf v^T} \\
  -{\bf v}  &  {\bf I} \\
\end{array}
\right)
\label{metric}
\end{equation}
where $v^2= \vert {\bf v} \vert^2$, ${\bf v^T}$ is the transpose of ${\bf v}$, and ${\bf I}$ indicates the identity matrix. The coefficients depend on the flow and, through the sound speed, on the fluid density. Hence, by tailoring the properties of the background flow and density it is possible to simulate a number of geometries of interest and related phenomena in general relativity. For example, in any region where the fluid is rotating at supersonic speed, no acoustic observers can remain at rest relative to an observer at infinity due to the supersonic dragging of inertial frames. A similar situation is found within the ergospheres surrounding rotating black holes. An acoustic horizon forms where the normal component of the flow passes from subsonic to supersonic, as any sound waves will be swept by the flow and be trapped in the supersonic region. The acoustic horizon causally disconnects this region of spacetime from the exterior, in analogy to the event horizon that general relativity predicts for black holes. 

Many hydrodynamic experiments realizing different analogue spacetime geometries have demonstrated several semiclassical effects. Surface wave experiments in water tanks allowed for the observation of negative-frequency waves at the event horizon \cite{rousseaux2008,rousseaux2010}, stimulated Hawking emission \cite{weinfurtner2011} classical Hawking correlations \cite{euve2016} and wave scattering processes \cite{euve2020}. Experiments in Bose-Einstein condensates (BECs) are approaching the sensitivity required to detect quantum correlations between outgoing and incoming phonon pairs at the horizon \cite{steinhauer2016}, a key signature of the quantum nature of acoustic Hawking radiation. Evidence of entanglement between Hawking particles has also been claimed \cite{steinhauer2016}, although this result has been questioned in Ref. \cite{leo18} (see also author's reply in Ref. \cite{stein18}). The measured correlation spectrum of phonon fluctuations near the horizon is also consistent with a thermal spectrum, with a temperature determined by the surface gravity \cite{deNova}. In this regard, it is important to note that dispersive effects significantly modify the Hawking spectrum \cite{isoard20,jacq20}. Strictly speaking, there is no even notion of temperature because the flux becomes frequency dependent. On the other hand, an approximate thermality should be recovered in the long wavelength regime and a subsequent theoretical analysis substantiated the experimental determination of the Hawking temperature presented in Ref. \cite{deNova} (please refer to Refs. \cite{isoard20,isoardthesis} for an exhaustive discussion of this issue). Even the stationarity of the process has been recently probed \cite{kolobov}, in line with the predictions of Hawking's theory. Unruh radiation has been detected in BECs, where a parametric modulation of the system simulated an accelerating reference frame \cite{unruhanalog}. Rotating black holes in (2+1)-dimensions have been generated in surface-wave experiments \cite{torres,torres2020}, where the first evidence of rotational superradiance was provided. These black holes have also been created later in quantum fluids of light \cite{vocke2018}. Superradiance has also been observed for acoustic waves scattered by a rotating absorbing cylinder \cite{cromb} and, recently, for modes amplified at the ergoregion in a photon superfluid \cite{braidotti2022,braidottiavs}. Acoustic analogues of cosmological redshift and Hubble friction \cite{eckel}, and cosmological particle production have also been experimentally demonstrated \cite{glorieux2022}.

After more than a decade of successful experiments testing different aspects of classical and semiclassical gravity, it is natural to wonder to what extent the analogy can be pushed into the domain of quantum gravity. In particular, two of the deepest puzzles in physics are related to classical or semi-classical black holes, and a complete quantum theory is expected to potentially solve them. 

First, the emission of Hawking radiation at the horizon is predicted to gradually diminish a black hole's mass, ultimately resulting in its complete evaporation. 
The eventual outcome of this process is a thermal bath of particles in a flat spacetime which would only contain information regarding the mass, electric charge, and angular momentum of the original black hole. This results into the near-complete loss of information contained in the constituents forming the black hole's initial state. This is inconsistent with unitarity, a central feature of quantum evolution, which preserves the information in a quantum system undergoing the evolution.

Second, curvature singularities, where the concepts of space and time break down catastrophically, as in the final stage of black hole collapse, or at $t=0$ for an expanding universe, are perhaps the most dramatic consequences of classical general relativity. The physical reality of these spacetime singularities has long been debated, and there is broad consensus that a quantum theory describing gravity at short distances could remove them.

Drawing from recent laboratory and theoretical advances, we describe a variety of acoustic analogue-gravity scenarios in hydrodynamic systems which might shed light on both of these fundamental puzzles.

Turning first to black hole radiance and evaporation, it has been suggested that the information loss paradox might be naturally solved by a fully-consistent quantum theory of gravitation, where black holes would also be treated quantum-mechanically \cite{ko}. Interestingly, analogue gravity in quantum fluids at least heuristically supports this view. The evolution of the complex order parameter $\psi$ describing the nonlinear dynamics of a generic quantum fluid is of course unitary, as prescribed by quantum theory. As usual in non-relativistic quantum mechanics, we can associate to the wave function $\psi=\sqrt{\rho}\,e^{i/\hbar S}$ the probability density current ${\bf J}= (\hbar/m) \,{\rm Im} (\psi^* \nabla \psi)= \rho {\bf v}$, where ${\bf v}=(1/m)\nabla S$ is the flow (fluid velocity field) and $\rho=\vert \psi \vert^2$ the fluid density. 
If now we expand the order parameter as the mean-field (classical) background $\psi_0$, which defines the spacetime geometry through the metric coefficients ${\bf v_0}$ and $c_s(\rho_0)$, and the quantum phonon fluctuations $\psi_1$ propagating on it, such that $\psi=\psi_0 + \epsilon \psi_1 + \mathcal{O}(\epsilon^2)$, where $\epsilon \ll 1$ is a perturbation parameter, unitarity is generally not preserved. In the presence of an acoustic black hole (a region of the background flow where ${\bf v_0} > c_s$), the information about the quantum state of an infalling phonon crossing the horizon would cease to be accessible to any acoustic observer in the outer region, leading to an apparent loss of unitarity.  

This simplistic argument suggests that the paradox may arise from an incomplete knowledge of the system, i.e. from the fact that we are trying to describe the full evolution solely in terms of the classical spacetime geometry (the background flow) and its linear quantum excitations (the phonons). The missing information is encoded in the neglected $\mathcal{O}(\epsilon^2)$-nonlinear terms which take into account the backreaction, i.e. the effects of the phonons over the spacetime geometry $\psi_0$. In BECs, quantum correlations between the background flow microscopic degrees of freedom and the phonon excitations are expected to arise, leading to an entangled state \cite{liberati19}. Similarly, in the process of evaporation of a black hole we can hypothesize the existence of correlations between the Hawking quanta and the degrees of freedom of quantum spacetime originating from the matter fallen into the black hole, preserving the unitarity of the system as a whole \cite{perez}. Experiments in BECs are approaching the sensitivity required to detect quantum correlations between outgoing and incoming phonon pairs at the horizon \cite{steinhauer2016}. Advanced versions of these setups might soon be able to extract correlations between condensed atomic states (which, in the mean field limit, give rise to the global geometry), and their collective quantum excitations, thus paving the way for exploring the physics of event horizons which are purely quantum in nature.

The above ideas relate to the possibility of storing information in a kind of gravitational memory and, in turn, to the breakdown beyond classical general relativity of the popular No-Hair Theorem. This theorem states that black holes can be entirely described by only three independent, externally observable, parameters: mass, electric charge, and angular momentum. Any additional characteristics or details regarding the composition of the collapsed object or the matter fallen into the horizon (i.e. the potential hair) would be unavoidably lost. While this is the case for classical black holes, the scenario could change if the gravitational field is quantized \cite{calmet}. Infalling quantum particles would perturb the geometry, encoding information about their state in its gravitational field, thus offering a potential solution to the information paradox \cite{calmet2,cheng}. As a result, despite being identical in classical theory, two black holes with same mass, charge and angular momentum would nonetheless show tiny differences in their gravitational fields at the quantum level. 

Discrete Quantum Hair, independent of geometry \cite{krausswilczek,colemanwilczek,krausspreskill}, might also manifest itself through purely quantum geometric phases that are invisible classically but which could be observable in long-range global Aharonov-Bohm type scattering experiments. Quantum hair in BECs could appear in the form of the aforementioned quantum correlations between the atomic degrees of freedom constituting the background flow, and its quantum fluctuations. Alternatively, it might be possible to directly search for Quantum Hair through long-range global scattering measurements that involve topological phonons (collective excitations bearing nonzero topological charge) and rotating black hole geometries imprinted into the condensate. This approach would be fundamentally unfeasible in astrophysical systems.

Both the physics of black hole radiation, and the existence of a curvature singularity in the final stages of black hole collapse share one feature, the need to consider modes of arbitrarily high frequency.

In the case of black-hole evaporation the process involves the physics of fields at arbitrarily high energies where classical theory starts to break down \cite{unruh76}. Due to extremely high gravitational redshift near the horizon, the outgoing Hawking particles are anticipated to stem from extremely high-frequency modes infinitesimally outside of the event horizon. This  prompts  the question of whether any new physics at the Planck scale could alter the Hawking spectrum or even halt the evaporation process \cite{pchen}.
In addition to the pursuit of a self-consistent theory of quantum gravity, a plethora of studies have employed a phenomenological approach to explore the potential implications of quantum gravitational effects. One such approach involves modifying the standard energy-momentum dispersion relation to incorporate higher-order corrections \cite{amelino98,garay-stf,amelino-doubly,magueijo}. These corrections break the Lorentz symmetry at high energies or small distances, often associated with the Planck scale. The possible implications of modified dispersion relations have been investigated in astrophysical observations (for a comprehensive review, see \cite{liberati-test}) and in the thermodynamics of black holes \cite{amelino2004,ling,nozari}.

In general, imposing a discreteness in spacetime at the Planck scale seems incompatible with Lorentz invariance, because different inertial observers would not agree on such a minimal distance (although some candidate theories still preserve it, see e.g. \cite{speziale}, and the phenomenon of duality in string theories also provides an effective Lorentz invariant lower bound on distances). On the other hand, the interesting idea that Lorentz invariance, and the very concept of space and time, with its curvature reflecting a gravitational field, could be the result of collective (large-scale) phenomena emergent from the dynamics of some elementary quantum constituents has been advanced in the recent years \cite{hu,bombelli,konopka}. In this sense gravity would be more similar to thermodynamics -- i.e. a macroscopic statistical behaviour emergent from some underlying complex many-particle physics -- rather than to electromagnetism as a fundamental force. Interesting approaches in this sense include causal sets (collections of discrete spacetime points related by a partial order \cite{bombelli}), or quantum \emph{graphity} models \cite{konopka}, where points in spacetime are represented by nodes on a graph, connected by links that can be on or off.

A spacetime geometry that gradually emerges, along with an effective Lorentz symmetry, at larger scales is a common feature of classical and quantum fluids. At large scales or, equivalently, low phonon energies, the fluid (more precisely the classical expectation value of the mean fluid flow) provides an effective curved spacetime on which elementary excitations propagate. However, at microscopic scales, the description in terms of continuous variables no longer holds, and this underlying geometric structure (as well as the concept of collective excitations) becomes increasingly less well defined. The scale at which the higher-order corrections become effective and the Lorentz symmetry is broken plays the role of the Planck scale in quantum gravity phenomenology. In ideal fluids this corresponds to the mean intermolecular distance, in superfluids to the so-called coherence length and, generally, is associated with some fundamental scale of the underlying microscopic physics of the system.

Modified dispersion relations in quantum gravity models can be usually written as  
\begin{equation}
E^2 = m^2 c^4 + c^2 p^2\left(1 + \frac{p^2}{p_{P}^2} + ... \right)\ ,
\end{equation}
where $E$ and $p$ are the energy and momentum of the particle, $c$ is the light speed and $p_{P}$ is the Planck momentum (the momentum of a particle with de Broglie wavelength equal to the Planck length) \cite{slv}. 

Similarly, collective excitations in some atomic condensates \cite{silke-visser,eg2} (see also the recent experiment \cite{ferrari22}) and photon fluids \cite{marino2019} obey an energy-momentum dispersion relation of the form 
\begin{equation}
E^2 = m^2 c_s^4 + c_s^2 p^2\left(1 + \frac{p^2}{2 p_c^2}\right) \,\,,
\label{disp2}
\end{equation}
where $m$ is the rest mass of the excitation, $p$ its momentum and the speed of sound $c_s$ plays the role of the light speed. The critical momentum $p_c = M c_s$, that here plays the role of the Planck momentum, depends on the mass $M$ of the particles forming the condensate and is inversely proportional to the so-called healing length $\xi=\hbar/(M c_s)$ (the minimal length over which the condensate maintains its coherence), where $\hbar$ is the reduced Planck constant. 

The dispersion curve (\ref{disp2}) connects three distinct regimes based on the momentum of the fluctuations. When $p \gg p_c$, Eq.~(\ref{disp2}) approximates the free-particle behaviour $E \approx p^2/2 M$, where the energy of the excitations approaches the energy of the individual particles forming the background fluid. In the intermediate regime, $m c_s \lesssim p \ll p_c$, also known as the hydrodynamic limit, the excitations obey the relativistic energy-momentum relation $E \approx \sqrt{p^2 c_s^2 + m^2 c_s^4}$. When $m^2 > 0$, these excitations resemble phonons in classical fluids, but possess a finite rest mass (attractive interactions in quantum fluids yield $m^2 < 0$, i.e. tachyonic excitations with real energy and momentum and imaginary rest mass). For $p \ll m c_s$, the energy-momentum relation simplifies to $E \approx p^2/2m + m c_s^2$, where the first term is the kinetic energy of a particle with mass $m$, and the second term denotes its constant rest mass energy. This corresponds to the non-relativistic limit of the excitations' dynamics.

The length $\xi$ (or equivalently the mass $M$ or the momentum $p_c$) provides the scale at which Lorentz invariance is explicitly broken. At low momenta $p \ll p_c$ massless phonons propagate at the invariant universal speed $c_s$. Generalized phonon dispersion relations and their role in the black hole evaporation process have been discussed by Jacobson \cite{jacobson}, Unruh \cite{unruh95} and Corley and Jacobson \cite{Corley}. The analysis presented in these works supports the viewpoint that Planck-scale modes would not prevent the existence of black-hole radiation. However, they could have a notable impact on the emitted spectrum \cite{jacob93}.

The thermal nature of acoustic Hawking radiation, at least in the low-energy regime of Lorentz-invariant phonon dispersion, has been confirmed experimentally \cite{deNova}. On the other hand, most experiments so far have been designed to operate within the hydrodynamic limit, precisely to minimise the effects of high-energy modes and achieve a clear demonstration of the Hawking process. By "hydrodynamic limit," here we mean configurations where the dispersion characteristic lenght (healing length in BECs) is significantly smaller than the characteristic length scale in which the flow passes from subsonic to supersonic, i.e. of the
surface gravity scale, which fixes the Hawking temperature. In this regime the Hawking spectrum is expected to approach thermality at low wavenumebers \cite{isoard20}.
Next-generation experiments could instead engineer acoustic black holes in a regime where high-energy phonons could leave their signatures on the radiation spectrum in the form of tiny deviations from thermality \cite{PhysRevD.71.024028}. It is worth remarking that it is the perfect thermality of the Hawking spectrum, which involves the transition from a pure initial state to a mixed final state, that underlies the information paradox.

As previously discussed, in acoustic analogue gravity, the spacetime curvature, acting as the gravitational field, governs the propagation of phonons. At the level of the linear evolution of non-interacting excitations, all effects due to "gravitational" backreaction are neglected: i.e. the acoustic spacetime geometry is not modified by the perturbations propagating on it. On the other hand, backreaction effects are inherent to nonlinear acoustics, where any density perturbation induces a disturbance in the background flow. This disturbance, in turn, defines the effective spacetime geometry through which the waves propagate. Under specific conditions, this self-interaction can be described in terms of a curved spacetime which is generated by the wave itself and determines its propagation \cite{goulart1,goulart2,cherubini,marino2016,fisher-grav}.
This extends the analogue gravity approach, as it encodes in a geometric framework the dynamical interplay between the phonons and the acoustic metric. As we will discuss, these systems can also be exploited to address the problem of curvature singularities. 


Although the emergence of spacetime geometry and effective Lorentz invariance for sound is remarkable, and despite the geometric descriptions mentioned above, an aspect in acoustic analogues that is generally lacking is the emergence of a relativistic gravitational dynamics. A notable exception is provided by relativistic BEC models where the full dynamics of the system can be described in terms of the Nordstr\"{o}m scalar theory of gravitation \cite{belenchia14}. The spacetime curvature is sourced by the expectation value of the stress-energy of the quasiparticles, while the Newton and cosmological constants are related to the characteristic scales of the microscopic system (for a more comprehensive discussion on "emergent gravity" in fluids see \cite{nrp}). 

This is the first example of analogue gravity in which a Lorentz invariant, geometric theory of semiclassical gravity emerges from an underlying quantum theory of matter. While the aforementioned theory is solely a scalar theory of gravity, like general relativity it satisfies the strong equivalence principle \cite{deruelle}. On the basis of this result, it is possible that relativistic condensates with more complex interactions (see e.g. \cite{eg2,marino2019} where the emergence of a Newtonian-like gravitational dynamics have been demonstrated in the non-relativistic case), may provide testbeds for something even closer to an emergent Einstein-like theory.  

Relativistic BECs could in principle also illuminate issues associated with holography. Planar AdS black-hole solutions can be exactly mimicked using these systems in ($2+1$) dimensions \cite{anti1,anti4}. It would be then interesting to simulate such geometries, where one could directly probe the relation between the emergent Nordstr\"{o}m gravity and the quantum hydrodynamic degrees of freedom \cite{dey}.

An alternative example in which geometry emerges from quantum theory is the fuzzball proposal where the bulk geometry of a black hole is replaced by a sphere of strings with a definite volume \cite{fuzz, fuzz0}. Even though the fuzzball is an quantum object, a black hole horizon arises as an effective geometric structure \cite{fuzz1}. Interestingly, fuzzballs admit as stable solutions spacetimes with an ergosphere and no horizon \cite{fuzz12}, a configuration that is linearly unstable in classical gravitational theories  \cite{friedman,Chirenti:2008pf}. Quantized vortices in superfluid provide natural examples of these spacetime geometries, thus serving as valuable probes of possible emergent phenomena \cite{braidotti2022,braidottiavs}.

In fluid mechanics, as in many other branches of physics, singularities emerge at scales where a given model fails to fully describe the system, and novel physical effects become significant. Although the mechanisms driving singularity formation in fluids and Einstein gravity differ, examining how hydrodynamic singularities are regularized when the curvature of the acoustic metric is offset by quantum effects could offer insights into modelling similar phenomena in gravity theories. In the following we discuss an example of an acoustic curvature singularity, showing that it is naturally removed when quantum effects are taken into account (see also Ref. \cite{nrp}).

In classical inviscid fluids, a finite-amplitude density perturbation cannot propagate indefinitely because the flow develops a discontinuity in a finite time. Different points along the wave profile travel at different velocities, resulting in the self-steepening of the wavefront. Eventually, the wavefront's gradient will exceed the vertical, becoming infinitely steep and leading to an unphysical solution to the fluid equations, thus exhibiting the so-called gradient catastrophe \cite{landau}. Theoretical investigations of this problem date back to Riemann's pioneering work on discontinuous flows in 1860 \cite{riemann}. 

Within the analogue-gravity formalism the above self-interaction dynamics can be described geometrically in terms of an emergent acoustic metric, with the gradient catastrophe corresponding to the formation of a curvature singularity. Owing to the nonlinear coupling between density and fluid velocity, changes in the density wave profile continuously modify the underlying flow and thus the associated spacetime geometry. This, in turn, affects the density wave propagating on it.
 
Unlike the linear case, where the acoustic geometry is fixed and shaped by suitable external forces, here a dynamic curved spacetime emerges spontaneously, as a result of the backreaction of a density perturbation over the background flow. In this way an initially negligible curvature/steepening will progressively amplify over time without limit \cite{marino2016}. A direct connection between the gradient catastrophe and the curvature singularity of the emergent metric has been explicitly demonstrated through computations of appropriate curvature invariants \cite{marino2016}. It's worth noting, however, that this singularity is ``naked,'' meaning it is not hidden to external observers by an event horizon.

In the framework of BECs and superfluids, the presence of quantum effects, stemming from the microscopic structure of these systems, fundamentally alters the scenario. The condensate demonstrates a degree of rigidity against spatial variations of its density, given by the so-called Bohm quantum potential \cite{bohmqp}, $V_Q = (\hbar^2/2 m)\nabla^2 \rho^{1/2} / \rho^{1/2}$. This term is purely quantum in nature, since it goes to zero in the classical limit as the de Broglie wavelength goes to zero. It also depends on the curvature of the amplitude of the wave function. In (1+1) dimensions, up to proportionality factors, this quantity corresponds to the Ricci scalar $\mathcal{R}$ of the emergent acoustic metric generated by the wave, so that $\mathcal{R} \sim V_q$.

The quantum stiffness counterbalances the curvature preventing the formation of a singularity. 
At a critical point where the gradient catastophe would otherwise take place, rapid spatial oscillations of density and velocity develop \cite{fischer-cens} leading to the formation of dispersive shock waves. These can be seen as spacetime structures of maximal, though finite, curvature. The formation of dispersive shock waves is the subject of ongoing experiments in BECs \cite{dutton} and photon superfluids \cite{wan,xu,bienaime,Bendahmane}. From an analogue gravity perspective, then, these quantum fluids provide an interesting arena for exploring how quantum effects might remove singularities that would otherwise be present in classical systems.

\vspace{.2 in}

Analogue gravity systems have been demonstrated to provide exciting and useful testing grounds for classical and semi-classical general relativity. As we have described here, new laboratory techniques in classical and quantum hydrodynamics are providing opportunities to expand the reservoir of experimental analogue gravity candidates. At the same time, recent work aimed at unraveling a full quantum theory of gravity suggests numerous possible effects that cannot be tested directly in astrophysical or cosmological contexts. Fluid experiments may usefully allow us to explore phenomena where quantum effects, going beyond a classical analogue gravity context, may provide insights into the black hole information paradox and the possible avoidance of classical spacetime singularities in a quantum theory of gravity, as well as other possible quantum corrections to classical gravity. While the dynamics of these system are not precisely the same as for real gravitational systems, exploring how such phenomena as back reaction impacts on their evolution may provide important clues to investigating similar problems gravitational physics. These exciting possibilities add motivation to continue further exploring analogue gravity with the hope of providing new empirical guidance into developing a consistent theory of quantum gravity.

\end{document}